# On improving the accuracy of nonhomogeneous shear modulus identification in incompressible elasticity using the virtual fields method

Yue Mei[1,2,3], and Stephane Avril[3*]

[1]State Key Laboratory of Structural Analysis for Industrial Equipment, Department of Engineering Mechanics, Dalian University of Technology, Dalian 116023, P.R. China

[2]International Research Center for Computational Mechanics, Dalian University of Technology, P.R.China

[3]Mines Saint-Étienne, Univ Lyon, Univ Jean Monnet, INSERM, U 1059 Sainbiose, Centre CIS Saint-Étienne, France

Corresponding author*: avril@emse.fr

**Abstract**

This paper discusses an important issue about the virtual fields method when it is used to identify nonhomogeneous shear moduli of nearly incompressible solids. From simulated examples, we observed that conventional virtual fields, which assign null displacements on the entire boundary, do not perform well on nonhomogeneous and nearly incompressible solids. Thus, these conventional virtual fields should not be used for such materials. We propose two novel types of virtual fields derived from either finite element analyses performed on the same domain with homogeneous properties or computing the curl of a potential vector field. From a variety of simulated and experimental examples, we observe that the proposed virtual fields significantly improve the accuracy of the estimated shear moduli of nonhomogeneous and nearly incompressible solids. Furthermore, the sensitivity to noise of the proposed approach is moderate and the approach can handle cases with unknown boundary conditions. Therefore, based on this careful and thorough analysis, it is concluded that the proposed approaches are a significant improvement of the VFM to identify nonhomogeneous shear moduli in nearly incompressible solids.

**Introduction**

Full-field measurement techniques have permitted significant progress in the identification of mechanical properties of solids [1]. To achieve this, numerous inverse algorithms have been proposed to identify mechanical properties from measured displacement fields across the domain of interest. Generally, the inverse algorithms can be categorized into two types: iterative and direct inversion methods. In iterative inversion methods, the inverse problem is posed to be a constrained optimization problem where the full-field displacement [2-6] or the resulting stress fields [7-10] are minimized. For reconstructing the nonhomogeneous mechanical property distribution of solids, a regularization term is usually introduced to avoid the over-fitting issue in reconstructing nonhomogeneous mechanical property distribution [6, 12]. This method is highly robust, insensitive to the noise level and can be easily generalized to nonlinear mechanical property identification [11]. However, this method is computationally costly and requires loading conditions which in some cases are unknown. The direct inversion method has been widely used in identifying homogenous and nonhomogeneous linear elastic properties [13-15]. The main idea behind is that the linear elastic properties can be expressed explicitly in terms of the strain and stress components, thus become the direct solution of the inverse problem. The direct inversion methods are very fast approaches to obtain the identified parameters or spatial variation of the linear elastic property distributions of solids. However, the direct inversion methods requires very high-resolution displacement fields restricting its application.

In 1989, Grédiac proposed a novel inverse algorithm based on the principle of virtual work referred to as the virtual fields method (VFM)[16]. Compared to optimization based approaches [11, 17-19], the VFM does not require solving the parameter identification problem iteratively when reasonable virtual fields are selected, thus remarkably reducing the computational cost. Due to this advantage, the VFM has been widely applied in material identification including but not limited to linear elastic [20], hyperelastic [21, 22] and nonelastic [23, 24] constitutive properties of solids.

Many rubber-like materials and most of biological soft tissues are nearly incompressible and their hydrostatic pressure term is not related to deformation through constitutive laws [25]. Since the hydrostatic pressure is usually not measurable either, this leads to a difficulty in identifying the mechanical properties. For the optimization based approaches, the cost function can be established without the hydrostatic pressure [26]. Consequently, estimated mechanical properties of solids are not influenced by the incompressibility. However, for the VFM, both displacements and the hydrostatic pressure are required for parameter identification. To address this issue, a type of "special" virtual fields were developed with virtual displacements zeroed on the entire boundary. However this approach, referred to as "conventional" onwards, is successful if and only if the gradients of the hydrostatic pressure are null across the domain (see Eq. (5))[20]. This approach has been commonly used to identify mechanical parameters of incompressible materials when the VFM was adopted [20, 27-29].

However, the assumption on null gradients of the hydrostatic pressure is only held for homogeneous materials in certain situations such as uniaxial tensile tests. For nonhomogeneous materials, the assumption is absolutely not satisfied and this may affect the accuracy of estimated mechanical properties. To the best of our knowledge, no studies have been done to test how this assumption affects the results of identification of nonhomogeneous materials.

In this paper, we will test the feasibility of the conventional type of special virtual fields to identify the regional mechanical properties of nonhomogeneous solids and examine how this assumption affects the identified parameters in a variety of examples. More importantly, we will propose two novel types of virtual fields to resolve the difficulty of VFM in identifying mechanical properties of incompressible and nonhomogeneous solids. The first type of virtual fields can be obtained from finite element simulations, and the second is from calculating the curl of a potential vector field.

In this work, we consider nonhomogeneous, incompressible linear elastic solids in which every nonhomogeneous region is known as *a priori*. The outline of this paper is as follows: In the **Methods** Section, we briefly review the mathematical foundation of the VFM, discuss the conventional virtual fields and propose the novel types of virtual fields for incompressible and nonhomogeneous solids. In the **Results** Section, a number of simulated and experimental displacement datasets of nonhomogeneous and incompressible solids are utilized to comprehensively compare different virtual fields and their performance on the accuracy of the identified shear moduli. Then we discuss the results in the **Discussion** Section and conclude this paper in the **Conclusion** Section.

**Methods**

## (1) Virtual fields method with its application to incompressible materials

The virtual fields method (VFM) is based on the principle of virtual work, which may be written, for quasi-static conditions, such as,

$$-\int_{\Omega} \boldsymbol{\sigma} : \boldsymbol{\varepsilon}^* \, d\Omega + \int_{\partial \Omega_t} \mathbf{t} \cdot \mathbf{u}^* \, d\partial \Omega_t = 0 \tag{1}$$

Where $\boldsymbol{\sigma}$ is the actual Cauchy stress tensor across the domain of interest $\Omega$. It is related to the strains through constitutive equations, strains deriving from gradients of the measured displacement field $\mathbf{u}$. $\mathbf{t}$ are the tractions applied on a part of the boundary $\partial \Omega_t$. $\mathbf{u}^*$ is a kinematically admissible virtual displacement field, and $\boldsymbol{\varepsilon}^*$ is the associated virtual strain field. For nonhomogeneous solids such as the problem presented in Fig. 1, Eq. (1) can be rewritten as:

$$-\sum_{i=1}^{n} \int_{\Omega_i} \boldsymbol{\sigma} : \boldsymbol{\varepsilon}^* \, d\Omega + \int_{\partial \Omega_t} \mathbf{t} \cdot \mathbf{u}^* \, d\partial \Omega_t = 0 \tag{2}$$

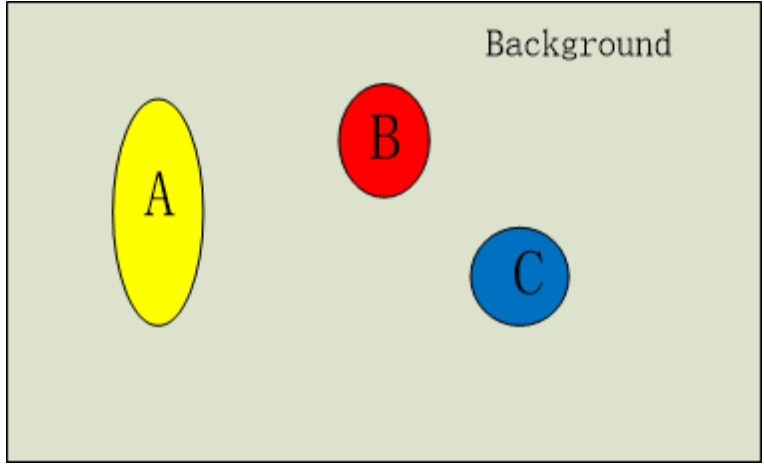

Figure 1: A nonhomogeneous problem domain. Regions A, B and C are inclusions with different shear moduli from the background.

where $\Omega_i$ is the i-th nonhomogeneous region (Background, Region A, B and C in Fig. 1). We assume homogeneous mechanical properties across each region $\Omega_i$. We also have the following relationship: $\bigcup_{i=1}^{n} \Omega_i = \Omega$ and $\Omega_i \cap \Omega_j = \emptyset$ when $i \neq j$. For an incompressible and linear elastic solid, Eq.(2) can be rewritten as:

$$-\sum_{i=1}^{n}\int_{\Omega_i}\left(2\mu_i\boldsymbol{\varepsilon}+p\mathbf{I}\right):\boldsymbol{\varepsilon}^*\mathrm{d}\Omega_i+\int_{\partial\Omega_t}\mathbf{t}\cdot\mathbf{u}^*\mathrm{d}\partial\Omega_t=0 \qquad (3)$$

where $\mu_i$ is the shear modulus value of $\Omega_i$ and $p$ is the hydrostatic pressure. To determine the shear moduli of $\Omega_i$, a total number of $n$ virtual fields or measured displacement fields should be applied to Eq.(3), leading to $n$ linearly independent equations. However, in practice, the hydrostatic pressure $p$ is highly difficult to measure. Thereby, it is necessary to make assumptions about the hydrostatic pressure values but if these assumptions are not satisfied, this will induce significant errors on the estimated mechanical properties of solids. To address this issue, a conventional choice of virtual fields which were previously considered [29, 30] was based on null displacements on the boundary

$$\mathbf{u}^*=0 \quad \text{on } \partial\Omega \qquad (4)$$

where $\partial\Omega$ is the boundary of the problem domain. In this case,

$$\int_{\Omega}p\mathbf{I}:\boldsymbol{\varepsilon}^*\mathrm{d}\Omega=\int_{\partial\Omega}p\mathbf{u}^*\mathrm{d}\partial\Omega-\int_{\Omega}\nabla p\cdot\mathbf{u}^*\mathrm{d}\Omega=-\int_{\Omega}\nabla p\cdot\mathbf{u}^*\mathrm{d}\Omega \qquad (5)$$

Assuming the hydrostatic pressure is homogeneous, Eq.(5) zeroes. However, this assumption only holds for homogeneous solids in very specific loading conditions. For nonhomogeneous solids, the conventional virtual fields will lead to an erroneous identification of shear moduli.

**(2) Novel types of virtual fields**

The solution to this problem is to define virtual fields satisfying:

$$\mathrm{div}\left(\mathbf{u}^*\right)=0 \text{ on } \Omega \qquad (6)$$

**Type 1: Virtual fields obtained from finite element simulations**

In the first type, we propose to reconstruct automatically virtual fields satisfying Eq.(6). For that, we are looking for two fields ($\mathbf{u}^*,p^*$) satisfying Eq. (6) and the following equation:

$$\nabla p^*+\mathrm{div}\left(\nabla\mathbf{u}^*+\left(\nabla\mathbf{u}^*\right)^\mathrm{T}\right)=\mathbf{0} \text{ on } \Omega \qquad (7)$$

Eq. (6) and Eq. (7) define the solution for an incompressible elastic problem in a homogeneous solid with a shear modulus equal to 1. The obtained displacement fields $\mathbf{u}^*$ automatically satisfy the incompressibility constraint. If we utilize them as virtual fields, the pressure term in Eq.(3) can be eliminated, that is,

$$\int_\Omega p\mathbf{I}:\boldsymbol{\varepsilon}^* d\Omega = 0 \qquad (8)$$

The problem consisting in Eqs. (6) and (7) is solved with the finite-element method for a set of appropriate boundary conditions. Examples of these boundary conditions will be given further.

When there are several inclusions with different material parameters as shown in Fig. 1 for instance, it is necessary to define as many virtual fields as there are inclusions (the number of unknowns is the number of inclusions, as the unknowns are the ratios between the modulus of each inclusion and the modulus of the background. Therefore it is necessary to solve Eqs (6) and (7) for as many different sets of boundary conditions as there are unknowns.

Finding all these virtual fields may sometimes be difficult. In that case we suggest to define a virtual field $\mathbf{u}_i^*$ for the identification of the modulus of each inclusion. We define then a volume of interest $\omega_i$ which is made of the background and of $\Omega_i$ only (then $\omega_i$ is $\Omega$ minus all the inclusions but $\Omega_i$). Then we find $\mathbf{u}_i^*$ by solving Eqs (6) and (7) on $\omega_i$ and by assigning $\mathbf{u}_i^*=0$ on the boundaries of all inclusions but $\Omega_i$. In the next section, we will compare the performance of conventional virtual fields with the virtual fields satisfying Eqs (6) and (7).

**Type 2: Virtual fields obtained from calculating the curl of a vector field**

We can also construct the virtual fields satisfying Eq. (6) by calculating the curl of a potential vector field since the divergence of a curl of a vector field is zero, that is,

$$\text{div}(\nabla \times \mathbf{F}) = 0 \qquad (9)$$

Where $\mathbf{F}$ is a nonzero vector field. Accordingly, we can define the virtual field as follows:

$$\mathbf{u}^* = \nabla \times \mathbf{F} \qquad (10)$$

However, since $\mathbf{u}^*$ should be kinematically admissible, it is not trivial to find the potential vector field $\mathbf{F}$ such that the displacement boundary conditions are satisfied. For some simple geometries such as square or ring models, as shown in Fig. 3, finding the potential vector field is rather simple. The general approach for derivation of all type 2 virtual fields utilized in this paper is presented in **appendix A**. Additionally, we should note that conventional virtual fields do not satisfy Eq. (9). This point will be discussed in the **Discussion** Section.

A number of numerical and experimental examples about the estimation of shear moduli in nonhomogeneous and incompressible solids with different problem domains will be considered. For the numerical examples, we follow the testing procedure shown in Fig. 2

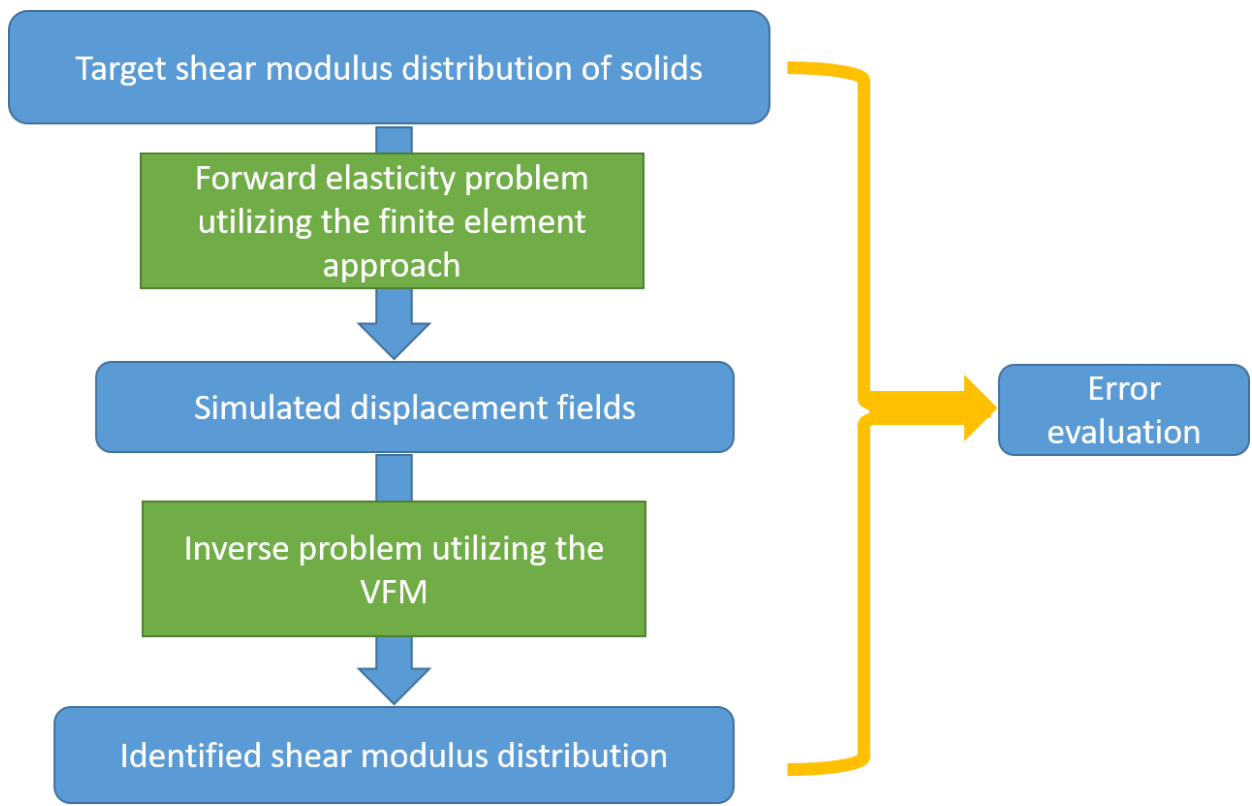

Figure 2: The flow chart of numerical testing for the inverse approaches.

**Results**

(1) Comparison of different virtual fields

In this subsection, a comparative study of different virtual fields to estimate the mechanical properties of solids is performed in the state of 2D plane strain. The domains of two examples are shown in Fig 3. In Fig 3(a), there is a stiff inclusion embedded in the soft background mimicking a tumor surrounded by the soft tissue [11, 31]. To acquire the simulated displacement fields, we solve the forward elasticity problem utilizing the finite element approach. We prescribe 1% compression on the top, restrict the vertical motion of the bottom edge, and fix the center node of the bottom edge. The problem domain is uniformly discretized by 101 nodes in each direction. The target shear modulus contrast between the stiff inclusion and soft background varies from 2.5 to 20. In solving the inverse problem, we merely estimate the shear modulus value in the inclusion assuming the shear modulus value of the background is known. The second example is an axisymmetric bi-layered ring structure representing the cross section of a carotid artery [32] as shown in Fig.3(b). The inner layer is stiffer than the outer layer. The shear modulus contrast of these two layers also varies from 2.5 to 20. The simulated data is obtained by solving a finite element problem where the pressure is applied on the inner wall. In the inverse problem, we assume the shear modulus value of the outer layer is known as a priori and identify the shear modulus value of the inner layer.

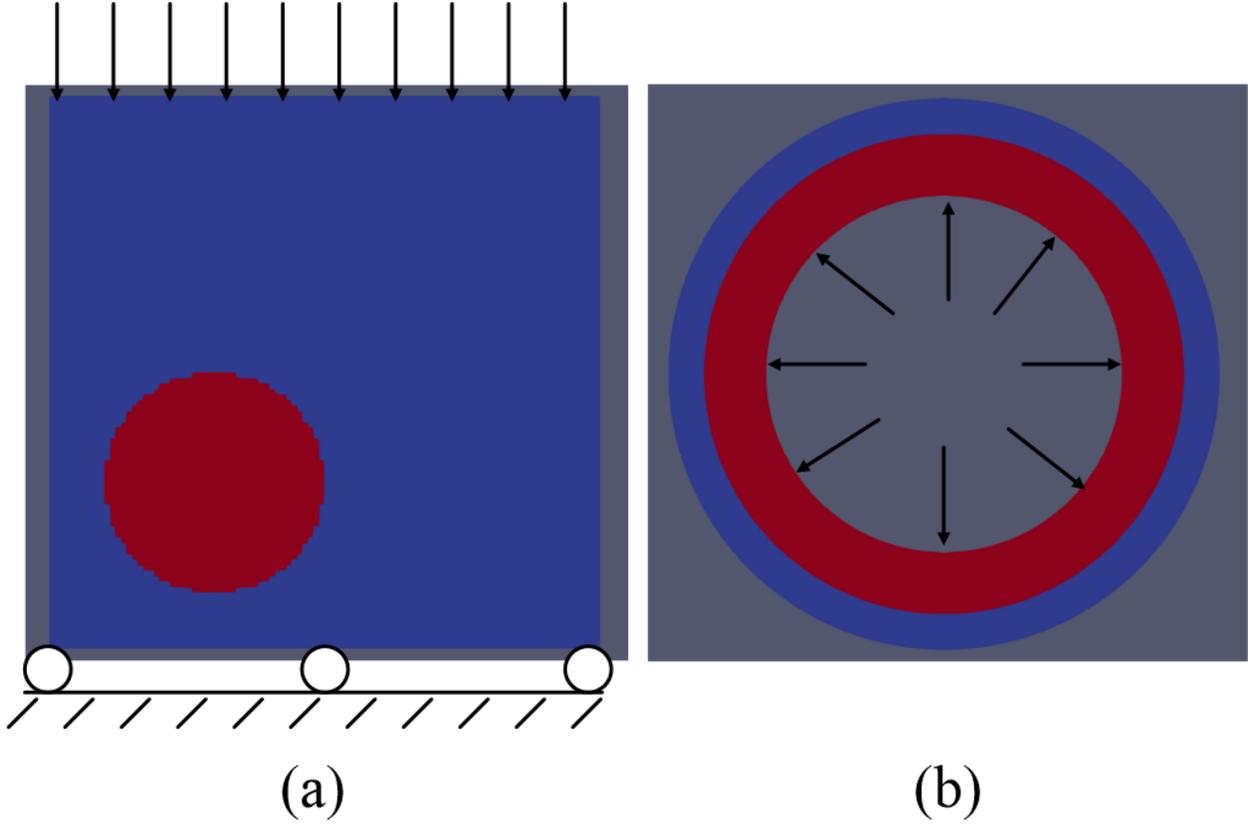

Figure 3: Problem domain of (a) the square model; (b) the ring model.

Four different virtual fields are used for comparison: (1) the conventional virtual fields that ensure null displacements on the entire boundary; (2) the actual displacement fields obtained from the resolution of the forward elasticity problem for the target shear modulus distribution; (3) Type 1 virtual fields: the displacement fields obtained by solving the forward elasticity problem with a homogeneous modulus using Eqs. (6) and (7); (4) Type 2 virtual fields that calculates curl of a vector field . For the domains shown in Fig 3, the conventional virtual fields may be written such as:

$$\begin{cases} u_x = 0 \\ u_y = (x-0)(x-L)(y-0)(y-L) \qquad \text{Square model} \\ u_z = 0 \end{cases}$$
$$\begin{cases} u_\theta = 0 \\ u_r = (r-r_0)(r-r_1) \qquad \text{Ring model} \\ u_z = 0 \end{cases} \qquad (11)$$

where $L$ is the side length of the square model and the origin of the coordinate system is set at the left bottom corner, $r_0$ and $r_1$ are the radii of the inner and outer walls, respectively.

Type 2 virtual fields is chosen as:

$$\begin{cases} u_x = x(L-2y) \\ u_y = (y-0)(y-L) \\ u_z = 0 \end{cases} \quad \text{Square model}$$

$$\begin{cases} u_\theta = -\left(3r^2 - 2(r_0+r_1)r + r_0 r_1\right)\theta \\ u_r = (r-r_0)(r-r_1) \\ u_z = 0 \end{cases} \quad \text{Ring model} \tag{12}$$

It is very interesting to see that for the ring model, though the radial displacement components are the same in Eqs. (11) and (12), the circumferential displacement component $u_\theta$ is nonzero for the type 2 virtual field which is not axisymmetric any more.

Fig 4(a) and (b) show the relative error between the estimated and exact shear modulus values of the inclusion for the square model and the inner layer for the ring model. The relative error is defined as

$$error = \left|\frac{\mu - \bar{\mu}}{\bar{\mu}}\right| \times 100\% \tag{13}$$

We observe that the virtual fields (1) induce over 20% relative error in estimated shear moduli and the relative error increases with the increase of the target shear modulus ratio. This is due to the fact that a higher target shear modulus ratio increases the heterogeneity of the pressure across the domain. As a result, $\nabla p$ in Eq. (5) becomes larger leading to the increase of the relative error. Furthermore, the proposed virtual fields (3) and (4) perform remarkably well in estimating shear moduli, with relative error close to zero. Interestingly, for the square model, the relative error for the virtual fields (2) is significantly higher than that for the virtual fields (3) and (4).

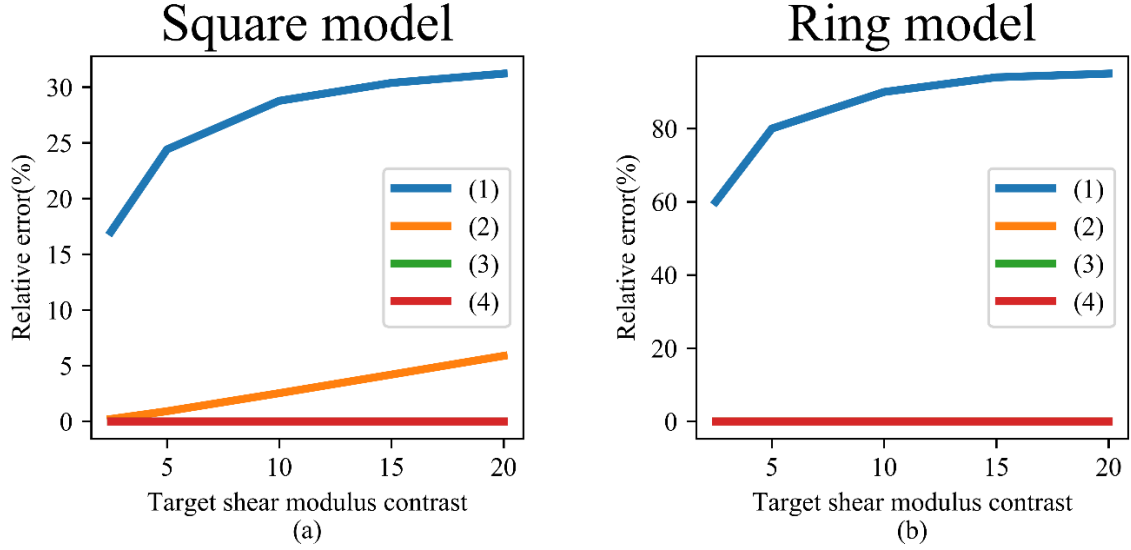

Figure 4: The relative error with varying target shear modulus contras. (a) Square model (the curves for virtual fields (3) and (4) are almost coincident); (b) ring model (the curves for virtual fields (2) ,(3) and (4) are almost coincident).

(2) Noise sensitivity for the proposed method.

For the proposed virtual fields, we add noise into the simulated displacements and study the noise sensitivity of the proposed approach. The noise level is defined as:

$$\text{noise level} = \frac{\sum_{i=1}^{N_{node}} (\bar{u}_i - u_i)^2}{\sum_{i=1}^{N_{node}} \bar{u}_i^2} \times 100\% \tag{14}$$

Where $N_{node}$ is the total number of nodes in the domain. $u_i$ and $\bar{u}_i$ are the noisy and exact nodal displacements, respectively.

The results of parameter estimation demonstrate that even 20% random noise in the displacements induces roughly 13% relative error in the estimated shear moduli for both novel virtual fields. We also observe that the relative error for the solids with higher target shear modulus ratio is generally higher than that with lower target shear modulus ratio. This is probably induced by the fact that the resulting displacements across the inclusion for the solids with higher target shear modulus ratio is smaller, hence more sensitive to noise. Overall, the examples presented in this subsection reveal that the proposed approaches have a low sensitivity to noise.

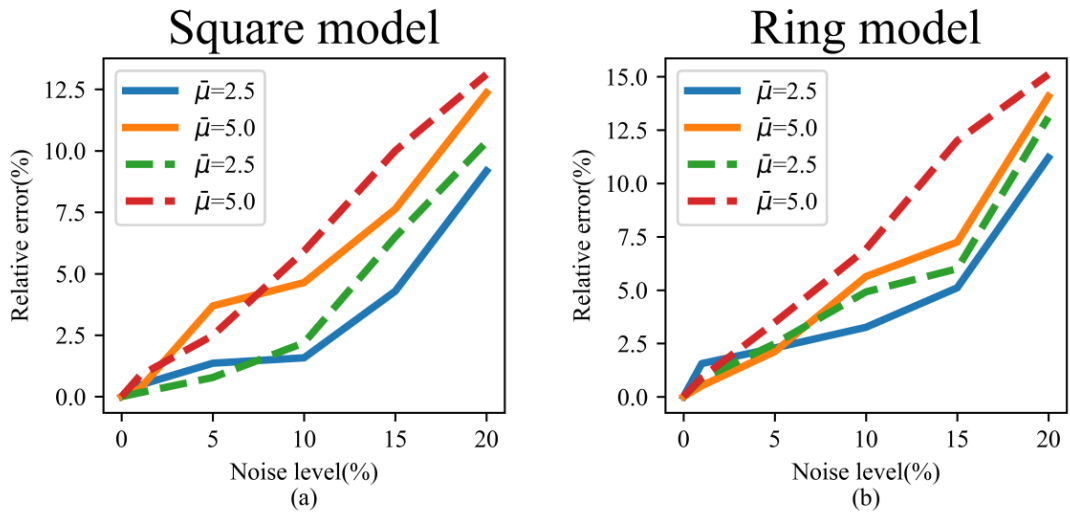

Figure 5: The relative error for two different target shear modulus ratios $\bar{\mu} = 2.5$ and 5 with respect to varying noise level (a) Square model; (b) ring model. Solid line: the virtual fields (3); dash line: the virtual fields (4)

(3) Multiple inclusion cases

We also tested the feasibility of the proposed method to estimate shear moduli for more than two regions. In this case, we do not assume the shear moduli of the background for the square models in Fig 6(a),(b) and the inner layer for the ring model in Fig. 6(c). Thereby, three shear moduli for each case have to be determined. As discussed in the **Methods** Section, a total number of 3 virtual fields and measured displacement fields should be utilized to determine all the unknown mechanical properties. Thus, we compare the relative error when we use 3 virtual fields for both new types of virtual fields or 3 measured displacement fields for the type 1 virtual fields. The three virtual fields for type 1 virtual fields are obtained from solving elasticity problems for the homogeneous problem domain posed by Eqs (6) and (7) with different boundary conditions (see Fig 7). Similarly, the three "simulated" measured displacement fields are also acquired from solving elasticity problems for the problem domain with the target shear modulus distribution (see Fig 8). Here, we do not introduce any noise into the simulated displacement fields. Tables 1 and 2 indicate that Type 1 virtual fields perform well in estimating the shear modulus for the incompressible and nonhomogeneous solids utilizing either 3 virtual fields or the measured displacement. For type 2 virtual fields, we chose the following three virtual fields for the square model:

$$
\begin{aligned}
&1 \quad \begin{cases} u_x = x(L-2y) \\ u_y = (y-0)(y-L) \\ u_z = 0 \end{cases} \\
&2 \quad \begin{cases} u_x = x \\ u_y = -y \\ u_z = 0 \end{cases} \\
&3 \quad \begin{cases} u_x = x(x-L) \\ u_y = -y(2x-L) \\ u_z = 0 \end{cases}
\end{aligned}
\tag{15}
$$

Table 3 reveals that type 2 virtual fields also perform well in identifying the shear moduli for the incompressible and nonhomogeneous solids with high accuracy. Besides, comparing Tables 1 and 3, we observe that type 2 virtual yields even more accurate identified parameters.

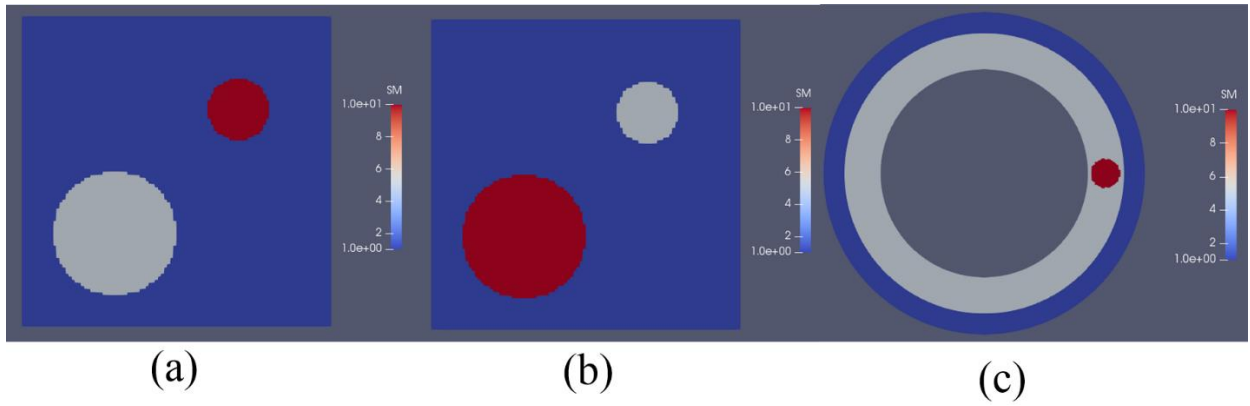

Figure 6: Problem domains for three different cases (a) The bigger inclusion is softer than the smaller inclusion; (b) The bigger inclusion is stiffer than the smaller inclusion;(c) A stiff inclusion is embedded in the inner layer of a bilayered cylinder mimicking an artery.

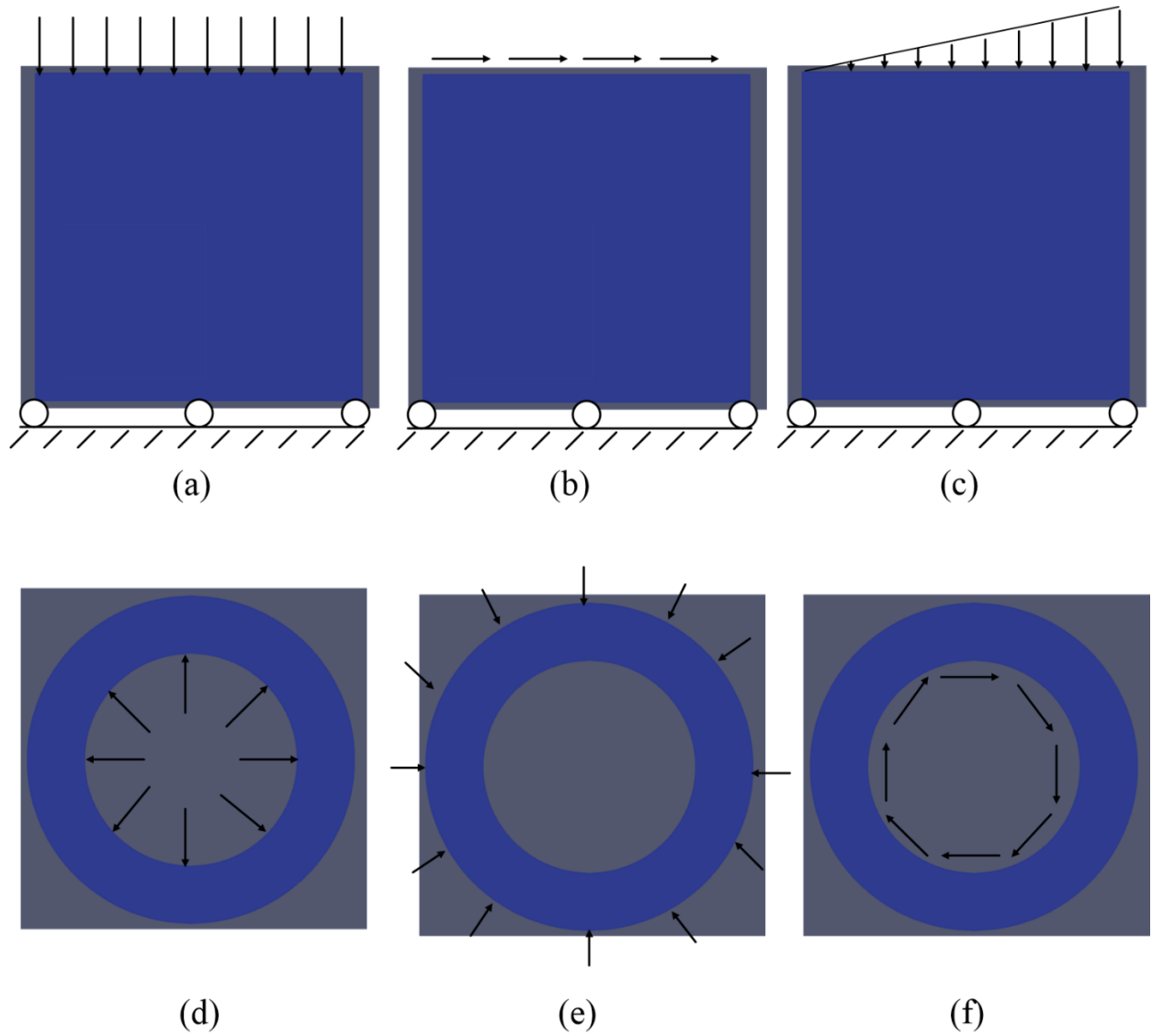

Figure 7: The first row (a, b and c) shows three different types of boundary conditions to establish three different virtual fields for the square model; The second row (d, e and f) shows three different types of boundary conditions to establish three different virtual fields for the ring model.

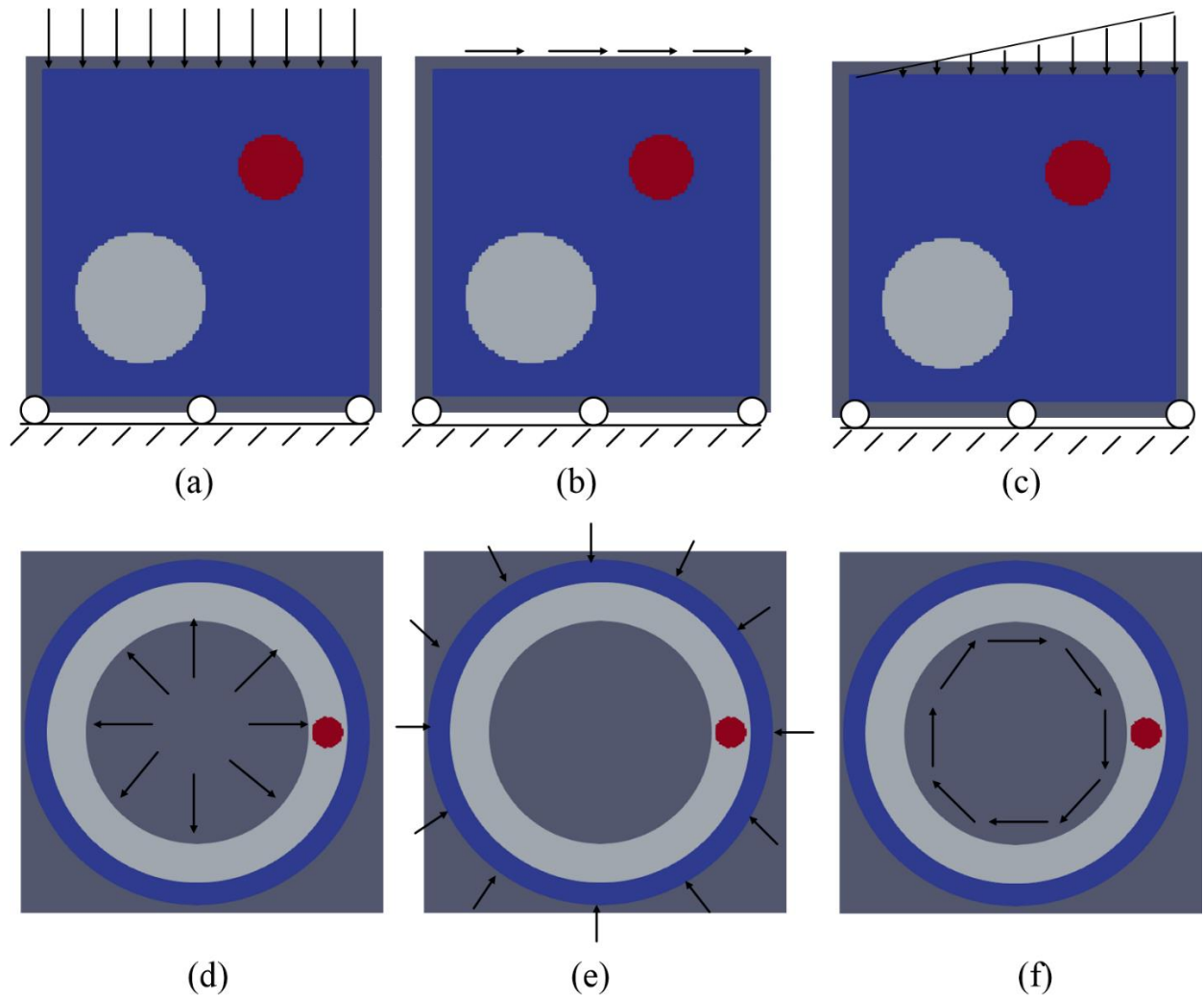

Figure 8: The first row (a, b, and c) shows three different loadings to simulate three different displacement field measurements for the square model; The second row presents three different loadings to simulate three different displacement field measurements for the ring model.

Table 1: Relative error between the estimated and target shear moduli for the square model using type 1 virtual fields.

|  |  | Background inclusion | Big inclusion | Small inclusion |
| --- | --- | --- | --- | --- |
| case (a) | 3 virtual fields | 0.01 | 0.08 | 0.00 |
|  | 3 experiments | 0.01 | 0.05 | 0.00 |
| case (b) | 3 virtual fields | 0.02 | 0.09 | 0.01 |
|  | 3 experiments | 0.00 | 0.00 | 0.00 |

Table 2: Relative error between the estimated and target shear moduli for the ring model using type 1 virtual fields.

|  |  | Inner layer | Outer layer | Inclusion |
|---|---|---:|---:|---:|
| case (c) | 3 virtual fields | 0.03 | 0.05 | 0.01 |
|  | 3 experiments | 0.07 | 0.03 | 0.02 |

Table 3: Relative error between the estimated and target shear moduli for the square model using type 2 virtual fields.

|  |  |  |  |  |
|---|---|---|---|---|
|  |  |  |  |  |
|  |  |  |  |  |

(4) Unknown applied loadings

In many engineering situations, the distribution of loads is unknown and very difficult to measure compared to the displacements. Such situations introduce another difficulty to solve the inverse problem related to the identification of material properties across the solids of interest. To this end, we also need to study the feasibility of the proposed virtual fields in cases where the loadings (non-zero tractions or forces) are unknown. We still take the uniaxial compression case presented in Fig 3(a) as an example and assume the traction applied on the top edge is unknown. In this case, we have to zero the vertical motion of the top edge in order to neglect the virtual work done by the unknown traction. To establish such type 1 virtual field, we assign a transverse virtual motion on the top edge and solve the incompressible elasticity problem posed by Eqs (6) and (7). We can also choose Eq. (12) as the type 2 virtual field to solve the parameter identification problem. Fig. 9 demonstrates that the proposed approaches are capable of successfully identifying the target shear modulus contrast between the inclusion and background. On the contrary, the conventional virtual fields (1) induce more than 15% error even when no noise is introduced into the measured displacements. Additionally, we also observe that type 2 virtual fields induce slightly less error compared to type 1 virtual fields.

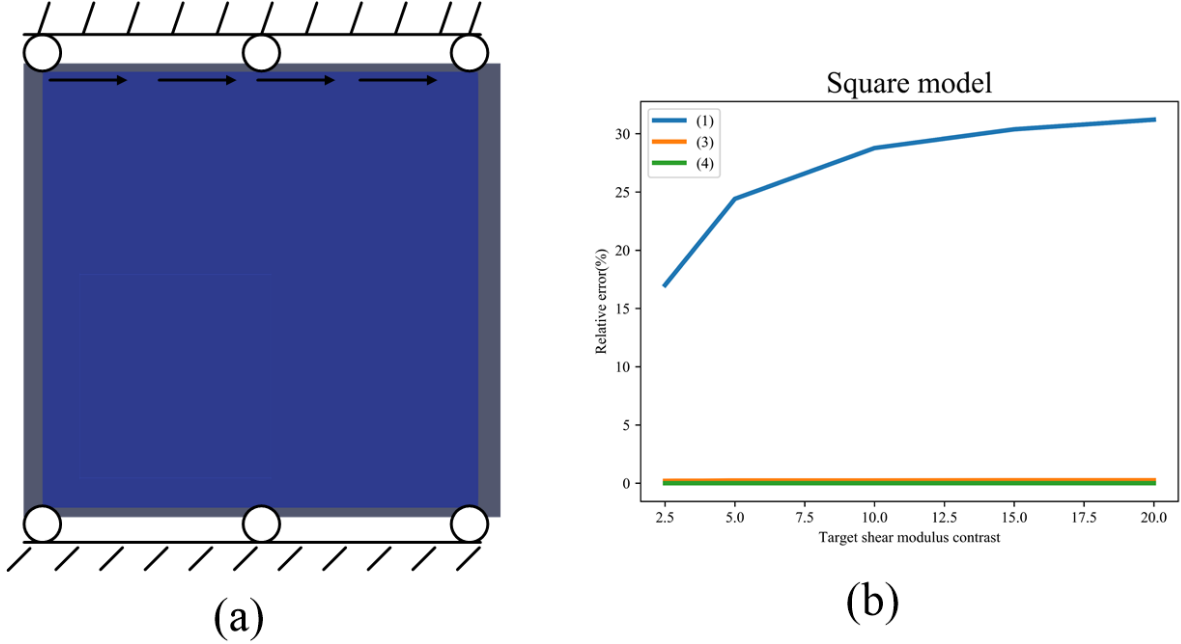

Figure 9: (a) Problem domain and the loadings for type 1 virtual fields; (b) Relative error between the estimated and target shear moduli utilizing virtual fields (1), (3) and (4).

(5) Experimental results

In this subsection, we test the proposed methods onto experimental data obtained on a cuboidal incompressible tissue-mimicking phantom where a stiff ellipsoid is embedded, as shown in [33]. The shear modulus ratio between the ellipsoid and the surrounded soft background is 4.0. The measured displacement fields of the sample subjected to uniaxial compression were acquired by Magnetic Resonance Imaging (MRI). More details on the experimental setup and data acquisition are presented in [27]. We assume the region of the stiff ellipsoid is determined and identify the shear modulus ratio between the ellipsoid and the background using the VFM. For virtual fields (1), the following functions are selected:

$$\begin{cases} u_x = 0 \\ u_y = (x - L_x/2)(x + L_x/2)(y - L_y/2)(y + L_y/2)(z - L_z/2)(z + L_z/2) \\ u_z = 0 \end{cases} \quad (16)$$

where $L_x$, $L_y$ and $L_z$ are the length, depth and height of the ellipsoid, respectively. For this conventional virtual fields, the recovered shear modulus ratio between the inclusion and background is 4.2. For the proposed type 1 and 2 virtual fields, the recovered shear modulus ratio is 4.05 and 4.04, respectively, very close to the target.

**Discussion**

This paper discusses a crucial issue on applying the VFM to identify nonhomogeneous mechanical properties in nearly incompressible solids and presents two novel types of virtual fields to address this issue. The proposed virtual fields are either acquired by solving the elastic incompressible problem with the finite element method for the corresponding homogeneous problem domain or calculating the curl from a potential vector field.

To simplify our analysis, we assumed that the solid studied herein is linear elastic, thus merely shear moduli should be identified in this paper. We performed a comparative study of the proposed and conventional virtual fields on identifying the regional shear moduli for incompressible elastic solids. We observed from both simulations and experiments that the proposed virtual fields performed much better in identifying the shear moduli of incompressible solids than the conventional virtual fields. This reveals that the conventional approach, though prevalent in solving parameter identification problems [20, 27-29], leads to errors in the identified shear modulus values. Conversely, the proposed virtual fields are capable of identifying shear moduli with very high precision for incompressible and nonhomogeneous solids. We also observed that the proposed approach is very moderately sensitive to noise, and this feature might be helpful when the displacement measurements are of low resolution. In some practical cases, the applied forces or tractions are unknown, and we also presented several examples to demonstrate that the proposed approach is capable of addressing this issue. To summarize, all the examples presented in the **Results** Section indicate that the proposed method is capable of identifying the mechanical properties with high accuracy, while the conventional approach induces significant error in the identified parameters.

Comparing both proposed virtual fields methods, type 1 virtual fields are more straightforward to acquire and can be easily generalized to any complex domains since finite element resolution of Eqs 6 and 7 are easily performed. On the contrary, due to the difficulty of finding the potential vector fields that satisfies the displacement boundary conditions, type 2 virtual fields are less easy to generalize. The merit of type 2 virtual fields is though that they are derived theoretically, thus no additional numerical error is introduced. That is the reason why the identified parameters by type 2 virtual fields are slightly more accurate than that of type 1 virtual fields (see Tables 1 and 3, and Fig. 9), despite the negligible level of the improvement. In future work, we should propose an approach to generalize type 2 virtual fields for complex domains.

For the conventional virtual fields, it does not satisfy Eq.(6) as shown in **Appendix B** where we take the square model (Fig. 3(a)) as an example. However, if the out of plane virtual strain component wrote $\varepsilon_{zz}^* = -\dfrac{\partial u_y^*}{\partial y}$ for the square model, it would satisfy Eq. (6). However, for the incompressible plane strain cases, the actual out of plane stress component would write $\sigma_{zz} = p$ which is usually not measurable. Thereby, even this modification of the conventional virtual fields would not work either.

To solve the identification problem with many different nonhomogeneous regions (see Subsection (3) in **Results** Section), we can either perform more experiments to acquire more measured datasets or employ different sets of boundary conditions to establish different virtual fields. Obviously, the latter is more practical since performing experiments demands more financial and time costs. Additionally, some experiments are highly difficult to perform such as the shear test for the ring model (see Fig 8(f)). Conversely, performing simulations to acquire virtual fields is more convenient.

We should also note that we restricted our analysis to linear elasticity. Nevertheless, the proposed approach can be easily applied to nonlinear and incompressible elastic solids [22, 27, 28]. For nonlinear and incompressible elasticity, the same issue induced by the hydrostatic pressure will occur for the conventional virtual fields. As the proposed virtual fields in nonlinear elasticity still satisfy the incompressible condition, the hydrostatic pressure term in the principle of virtual work can be canceled out and the issues can be smoothly handled.

**Conclusion**

In this paper, we proposed novel approaches to establish virtual fields either from finite element analyses for the homogeneous problem domain or calculating curl of a vector field. The resulting virtual fields remarkably improve the accuracy of identified regional shear moduli for linear elastic and nonhomogeneous solids when the VFM is utilized. To test the feasibility of the proposed methods and compare it with the conventional virtual fields, a variety of simulated and experimental examples were presented. The results of these identification problems showed that the conventional approach induced significant errors in the identified shear moduli and the error increased with the increasing target shear modulus ratio. The proposed method was able to estimate the shear moduli with high accuracy even in the presence of high level of noise. This study clearly demonstrated that the conventional virtual fields which assign null displacements on the entire boundary should not be selected to estimate elastic moduli in nonhomogeneous and incompressible solids, while our proposed approaches should be systematically considered for such situations.

**Acknowledgement**

Yue Mei acknowledges the support from the Fundamental Research Funds for the Central Universities (Grant No. DUT19RC(3)017), Stephane Avril acknowledges the support from the European Research Council for grant ERC-2014-CoG BIOLOCHANICS. We also acknowledge the anonymous reviewer for providing the idea of type 2 virtual fields.

**Appendix A    Derivation of type 2 virtual fields used in this paper**

For the square model, all the type 2 virtual fields utilized in this paper has been presented in Eq. (15). To derive the first virtual field in Eq. (15), we assume the virtual displacement components with the following form:

$$\begin{aligned} u_x^* &= \frac{\partial F_z}{\partial y} - \frac{\partial F_y}{\partial z} = xf(y) \\ u_y^* &= \frac{\partial F_x}{\partial z} - \frac{\partial F_z}{\partial x} = y(y-L)g(x) \\ u_z^* &= \frac{\partial F_y}{\partial x} - \frac{\partial F_x}{\partial y} = 0 \end{aligned} \quad (17)$$

Where the vector field is written such as $\mathbf{F} = \begin{bmatrix} F_x & F_y & F_z \end{bmatrix}^T$. This displacement field satisfies the displacement boundary conditions. To simplify the problem, we set $F_z = 0$, thus leading to

$$\begin{aligned} F_x &= y(y-L)g(x)z + h_2(x,y) \\ F_y &= -xf(y)z + h_1(x,y) \end{aligned} \quad (18)$$

Substituting Eq. (18) into the third equation in Eqs.(17) yields:

$$-f(y)z+\frac{\partial}{\partial x}h_1(x,y)=(2y-L)g(x)z+\frac{\partial}{\partial y}h_2(x,y) \tag{19}$$

If we set $h_1(x,y)=h_2(x,y)=0$, then we obtain $f(y)=(L-2y)$ and $g(x)=1$. We skip the derivation of other virtual fields since the procedure is similar.

**Appendix B  Proof that the conventional virtual fields does not satisfy Eq. (6)**

Consider the square model as shown in Fig. 3(a), the conventional virtual fields are given as:

$$\begin{cases} u_x^* = \dfrac{\partial F_z}{\partial y} - \dfrac{\partial F_y}{\partial z} = 0 \\[4pt] u_y^* = \dfrac{\partial F_x}{\partial z} - \dfrac{\partial F_z}{\partial x} = (x-0)(x-L)(y-0)(y-L) \\[4pt] u_z^* = \dfrac{\partial F_y}{\partial x} - \dfrac{\partial F_x}{\partial y} = 0 \end{cases} \tag{20}$$

Thereby,

$$\frac{\partial F_z}{\partial y} = \frac{\partial F_y}{\partial z} \tag{21}$$

This leads to the following relationship by taking partial derivatives of Eq. (21) with respect to $x$

$$\frac{\partial^2 F_z}{\partial x \partial y} = \frac{\partial^2 F_y}{\partial x \partial z} \tag{22}$$

Taking advantage of Eq. (22), the out of plane normal strain component can be written as:

$$\varepsilon_{zz}^* = \frac{\partial u_z}{\partial z} = \frac{\partial^2 F_y}{\partial x \partial z} - \frac{\partial^2 F_x}{\partial z \partial y} = \frac{\partial^2 F_z}{\partial x \partial y} - \frac{\partial^2 F_x}{\partial z \partial y} = -\frac{\partial u_y^*}{\partial y} \tag{23}$$

Thus, to be a curl of a vector field, the out of plane strain component should be non-zero. Thus, the conventional virtual field Eq. (20) cannot be the curl of any vector field.